\documentclass[prd,aps]{revtex4}
\usepackage{amssymb,amsmath,amsfonts,amsbsy,epsfig}
\usepackage{slashed}
\usepackage{xcolor}
\newcommand{\be}{\begin{equation}}
\newcommand{\ee}{\end{equation}}
\newcommand{\bea}{\begin{eqnarray}}
\newcommand{\eea}{\end{eqnarray}}
\newcommand{\nn}{\nonumber}

\def\L{{\mathcal L}}

\def\D{{\mathcal D}}

\begin{document}

\title{Schwinger model \`a la Very Special Relativity}
\date{\today}

\author{
Jorge Alfaro$^{a}$, and Alex Soto$^{a}$
}

\affiliation{
$^{a}$Instituto de F\'{i}sica, Pontificia Universidad de Cat\'olica de Chile, \mbox{Av. Vicu\~na Mackenna 4860, Santiago, Chile}
}

\begin{abstract} 

In this work, we show that Lorentz invariant theories in $1+1$ dimensions admit new terms inspired by Very Special Relativity (VSR) theories. We have studied the Schwinger model in VSR. We show the axial current is classically conserved in the presence of a mass term coming from the VSR invariant terms but without standard Lorentz invariant mass. Furthermore, it is shown that both  the vector current as well as the axial current are modified with respect to the free case when the fermion is coupled to an external electromagnetic field due to the nonlocal operator present in the theory. The axial anomaly is computed, and we found the same standard topological invariant with a modification in the coefficient.

\end{abstract}

\maketitle 

\section{Introduction}



Quantum Electrodynamics in $1+1$ dimensions ($QED_2$) has been studied, and it has an exact solution discovered by Schwinger\cite{Schwinger:1962tp} when the fermion remains massless. In this model, called Schwinger Model, the photon acquires a mass $e^2/\pi$. This model has been studied and reviewed extensively (see for example \cite{Abdalla:1997qt,Abdalla:1991vua,Dittrich:1985tr,Manton:1985jm}) since this model presents confinement, because in two-dimensional space-time the Coulomb potential increases linearly, and it prevents the fermions to become free\cite{Wolf,Casher:1974vf}. Interesting properties as the nontrivial vacuum structure were studied in the work of Lowenstein and Swieca\cite{Lowenstein:1971fc}. Moreover, instantons in this model have been analyzed by Smilga\cite{Smilga:1993sn}. In addition, despite the massive Schwinger model is not exactly solvable, it has been reviewed too\cite{Coleman:1975pw, Coleman:1976uz}. Another essential feature in the Schwinger model is the presence of the chiral anomaly, which is easier to compute than in four dimensions. The axial vector current, which classically is conserved, gets a new term after radiative corrections. Thus,
\be
\label{stanomaly}
\partial_\mu j^{\mu5}=\frac{e}{2\pi}\epsilon^{\mu\nu}F_{\mu\nu}.
\ee
For  good reviews of this anomaly using a perturbative treatment see \cite{Peskin:1995ev,Harvey:2005it}.\\

Anomalies have not been studied yet in the context of $SIM(2)$ invariant theories. These theories began with the claim of Cohen and Glashow that nature could be described only with the Lorentz subgroup $SIM(2)$\cite{Cohen:2006ky}. This theory, studied in four dimensions and called Very Special Relativity (VSR), does not have invariant tensors, and it has the same important features of Special Relativity, like time dilation, velocity addition, and maximum attainable velocity. Under this framework, a fixed null vector $n$ transforms with a phase. Hence, new invariant terms like $n\cdot p/n\cdot q$ can be constructed in the lagrangian. As a consequence of this, the neutrino gets mass without new particles or violation of leptonic number\cite{Cohen:2006ir}. In four dimensions VSR has been studied in electrodynamics\cite{Cheon:2009zx}, and an important feature is the possibility and new consequences of a non-violating gauge invariant photon mass\cite{Alfaro:2019koq}. Also, the electroweak model under this formalism has been reviewed\cite{Alfaro:2015fha}. In addition, we found VSR studies in locally anisotropic cosmology\cite{Kouretsis:2008ha}, considering two-time physics\cite{Romero:2012vp}, curved space-time\cite{Muck:2008bd} and using gaugeon formalism\cite{Upadhyay:2016hdj}.\\

In this work, we will focus our attention on two dimensions instead of four, and we will analyze the chiral anomaly in the VSR-QED without the standard Lorentz invariant mass. The anomaly computation is easier than in four dimensions and it is simpler to test the new VSR-like terms here before we apply them in four-dimensional models. Meanwhile we have to keep in mind  that  in lower dimensions we have less degrees of freedom than in real life. In fact  there are not true dynamical degrees of freedom associated with the electromagnetic field in two dimensions\cite{Coleman:1976uz}.\\

Thus, the outline of this work is as follows. In section II we will analyze the Lorentz group for two dimensions and the connection with VSR theories. In section III we will review the classical conservation of the free vector and axial  current under the VSR formalism. In section IV we will derive the vector and axial  current when the fermion is coupled with an external electromagnetic field. Section V presents the path integral view of the lagrangian reviewed in section IV, and we will compute the vacuum polarization of the photon. In section VI, we will compute the expectation value of the vector current. In section VII, we present the result of the axial anomaly in the VSR two dimensional model, and finally, in section VIII, we will summarize and highlight the results.

\section{Lorentz Group in 1+1 dimensions}
We recall the Lorentz group as the group of transformations that left invariant the metric
\be
\label{deflor}
g^{\mu\nu}=g^{\rho\sigma}\Lambda^\mu_\rho\Lambda^\nu_\sigma.
\ee
In two dimensions we choose the metric as $g=diag(1, -1)$. If we consider an infinitesimal transformation $\Lambda^\mu_\rho=\delta^\mu_\rho+\omega^\mu_\rho$, it is easy to see that $\omega^{\mu\nu}=\omega^\mu_\rho g^{\rho\nu}$ is antisymmetric. Thus, due to the antisymmetry, the Lorentz group in $1+1$ dimensions has only one parameter. It means we have one generator of the Lorentz group. We define it as
\be
K=\begin{pmatrix}
    0 & 1 \\
   1 & 0
  \end{pmatrix}.
\ee
We notice that we can construct the Lorentz transformations applying successive transformations to the identity
\be
\Lambda(\theta)=\exp{(K\theta)},
\ee
where $\theta$ is a parameter. With this, the transformation reads

\be
\Lambda=\begin{pmatrix}
    \cosh\theta & \sinh\theta \\
   \sinh\theta & \cosh\theta
  \end{pmatrix}.
\ee
We can check $\Lambda$ effectively satisfies the identity (\ref{deflor}). Moreover, is easy to see there are not invariant tensors under this transformation. However, if we observe the following null vector
\be
n=\begin{pmatrix}
    1 \\
    1
  \end{pmatrix},
\ee 
it transforms with a phase under this transformation, $\Lambda n=e^\theta n$. Therefore, we can add to the lagrangian terms with fractions which contain $n$ as in the numerator as in the denominator, because they are invariant under the Lorentz transformation. This kind of terms have been studied in four dimensional VSR theories (see for instance \cite{Cohen:2006ir,Cheon:2009zx,Alfaro:2019koq,Alfaro:2015fha}), where the null vector $(1,0,0,1)$ transforms in the same way under $SIM(2)$ group transformations. Nevertheless, these terms have not been incorporated in two dimensional works in Lorentz invariant theories and they could be added.

\section{Classical axial and vector currents for free VSR fermions}
Since the two dimensional Lorentz theories admit terms with the null vector $n=(1,1)$ we proceed as in the VSR theories. The general VSR lagrangian for a free fermion is given by
\be
\L=\bar{\psi}\left(i\slashed{\partial}-M+i\frac{m^2}{2}\frac{\slashed{n}}{n\cdot \partial}\right)\psi.
\ee
Here, we use the slash notation $\slashed{n}=\gamma^\mu n_\mu$ and $\slashed{\partial}=\gamma^\mu \partial_\mu$. The $\gamma^\mu$ are the gamma matrices, which we have chosen in two dimensions as
\be
\gamma^0=\begin{pmatrix}
    0 & -i \\
   i & 0
  \end{pmatrix}, \qquad
\gamma^1=\begin{pmatrix}
    0 & i \\
   i & 0
  \end{pmatrix}.  
\ee
Also, we can define $\gamma^5=\gamma^0\gamma^1$, which will be useful later. In this representation, $\gamma^5$ is given by
\be
\gamma^5=\begin{pmatrix}
    1 & 0\\
   0 & -1
  \end{pmatrix}.
\ee

In addition $m$ is a mass parameter in the SIM(2) theory which parametrizes the deviations of the Lorentz Symmetry. Also, we can associate this parameter with the neutrino mass (see \cite{Cohen:2006ir}). We notice if $m\to0$, we recover the standard result. In addition, $M$ is the standard Lorentz invariant fermion mass. We will consider from here $M=0$ to study the Schwinger model in VSR. Nevertheless, under this assumption, our model is a fermion with a small mass $m$. Although there is not standard mass, our fermion is massive due to the VSR term. Thus, in VSR we have  a slightly different massive Schwinger model. Also, $\psi$
is a two component spinor. As $M=0$, the lagrangian is
\be
\label{lagfree}
\L_0=\bar{\psi}\left(i\slashed{\partial}+i\frac{m^2}{2}\frac{\slashed{n}}{n\cdot \partial}\right)\psi.
\ee
From the equation (\ref{lagfree}) we compute the equations of motion for $\bar{\psi}$ and $\psi$
\bea
\bar{\psi}\left(-i\overset{\leftarrow}{\slashed{\partial}}-i\frac{m^2}{2}\frac{\slashed{n}}{n\cdot \overset{\leftarrow}{\partial}}\right)&=&0 \label{eqbpsi}\\
\left(i\slashed{\partial}+i\frac{m^2}{2}\frac{\slashed{n}}{n\cdot \partial}\right)\psi&=&0 \label{eqpsi}
\eea
where $\overset{\leftarrow}{\partial}$ indicates the derivative acts over the object in the left. If we multiply by $\psi$ in the right in (\ref{eqbpsi}) an by $\bar{\psi}$ in the left in (\ref{eqpsi}) and we sum both we get
\be
\partial_\mu\left(\bar{\psi}\gamma^\mu\psi+\frac{m^2}{2}\left(\frac{1}{n\cdot\partial}\bar{\psi}\right)\slashed{n}n^\mu\left(\frac{1}{n\cdot\partial}\psi\right)\right)=0.
\ee
Therefore, we define the expression in the bracket as the free vector current
\be
\label{veccurrent}
j_{free}^\mu=\bar{\psi}\gamma^\mu\psi+\frac{m^2}{2}\left(\frac{1}{n\cdot\partial}\bar{\psi}\right)\slashed{n}n^\mu\left(\frac{1}{n\cdot\partial}\psi\right).
\ee
We can proceed in an analogue way to define the axial current $j^{\mu5}$ as
\be
\label{axcurrent}
j_{free}^{\mu5}=\bar{\psi}\gamma^\mu\gamma^5\psi+\frac{m^2}{2}\left(\frac{1}{n\cdot\partial}\bar{\psi}\right)\slashed{n}\gamma^5n^\mu\left(\frac{1}{n\cdot\partial}\psi\right).
\ee
We observe that $\partial_\mu j_{free}^\mu=0$ and $\partial_\mu j_{free}^{\mu5}=0$; thus, both currents are conserved at the classical level.\\

In two dimensions the relation $\gamma^\mu\gamma^5=-\epsilon^{\mu\nu}\gamma_\nu$ holds, where $\epsilon^{\mu\nu}$ is the two-dimensional Levi-Civita symbol. Hence, we apply this relation in the equation (\ref{axcurrent}) and we use the equation (\ref{veccurrent}) to get
\be
j_{free}^{\mu5}=-\epsilon^{\mu\nu}j^{free}_\nu+\frac{m^2}{2}\left(\frac{1}{n\cdot\partial}\bar{\psi}\right)(\epsilon^{\mu\nu}\slashed{n}n_\nu+\slashed{n}\gamma^5n^\mu)\left(\frac{1}{n\cdot\partial}\psi\right).
\ee
Since $n^0=n^1=1$ and $n_0=-n_1=1$ we found $\epsilon^{\mu\nu}\slashed{n}n_\nu+\slashed{n}\gamma^5n^\mu=0$. Thus
\be
\label{jrelation}
j_{free}^{\mu5}=-\epsilon^{\mu\nu}j_{\nu}^{free},
\ee
as in the standard case.

\section{Currents for fermions in an external electromagnetic field}
Let us consider the lagrangian for a fermion with only a VSR mass term coupled with an external electromagnetic field $A_\mu$.
\be
\L=\bar{\psi}\left(i\slashed{D}+i\frac{m^2}{2}\frac{\slashed{n}}{n\cdot D}\right)\psi,
\ee
where $D_\mu=\partial_\mu+i e A_\mu$. From this lagrangian we get the equations of motion for $\psi$ and $\bar{\psi}$
\bea
\left(i\slashed{D}+i\frac{m^2}{2}\frac{\slashed{n}}{n\cdot D}\right)\psi&=&0 \label{eomD1},\\
(D^{\dagger}_{\mu} \bar{\psi}) \gamma^{\mu} + \frac{1}{2} m^2 \left( \frac{1}{n
\cdot D^{\dag}} \bar{\psi} \right) \slashed{n} &=& 0 \label{eomD2},
\eea
where the operator $D^{\dagger}_{\mu}$ is defined as $D^{\dagger}_{\mu}=\partial_\mu-i e A_\mu$.\\

We proceed similarly as in the free case, multiplying by $\bar{\psi}$ in the left in (\ref{eomD1}), by $\psi$ on the right in (\ref{eomD2}) and sum both. It results in
\be
\label{cur1}
\partial_{\mu} (\bar{\psi} \gamma^{\mu} \psi) + \frac{1}{2} m^2 \left[
\bar{\psi} \slashed{n} \left( \frac{1}{n \cdot D} \psi \right) + \left(
\frac{1}{n \cdot D^{\dag}} \bar{\psi} \right) \slashed{n} \psi \right] = 0.
\ee

We can rewrite this expression as
\be
\partial_\mu\left[\bar{\psi} \gamma^{\mu} \psi+\frac{1}{2}m^2\left(\frac{1}{n \cdot D^{\dag}}\bar{\psi}\right)\slashed{n}n^\mu\left( \frac{1}{n \cdot D} \psi\right) \right]=0.
\ee

Thus, the current in this case is given by
\be
j^\mu=\bar{\psi} \gamma^{\mu} \psi+\frac{1}{2}m^2\left(\frac{1}{n \cdot D^{\dag}}\bar{\psi}\right)\slashed{n}n^\mu\left( \frac{1}{n \cdot D} \psi\right).
\ee

We notice the current for the fermion in the electromagnetic field is modified. Due to the non local term $(n\cdot D)^{-1}$ the current $j^\mu$ acquires a new dependence on $A$. We proceed in an analogue way to get $j^{\mu 5}$ and we get:
\be
j^{\mu 5}= \bar{\psi} \gamma^{\mu} \gamma^5\psi+\frac{1}{2}m^2\left(\frac{1}{n \cdot D^{\dag}}\bar{\psi}\right)\slashed{n} n^\mu\gamma^5\left( \frac{1}{n \cdot D} \psi\right)
\ee
Despite the modification of the current from its free counterpart, for the axial and vector currents the relation
\be
\label{jrelation2}
j^{\mu5}=-\epsilon^{\mu\nu}j_{\nu},
\ee
still holds, and they are classically conserved ($\partial_\mu j^\mu=\partial_\mu j^{\mu5}=0$).\\

Notice that both currents are gauge invariant under $\psi\to e^{-i e\chi}\psi $ and $A_\mu\to A_\mu+\partial_\mu\chi$. Then we can work in the gauge $n\cdot A=0$. In this gauge both currents reduce to the free case currents.

\section{Path Integral, Feynman Rules and Photon Self-Energy}
In this section we will analyze the fermion coupled to an external electromagnetic field using the path integral formalism. The generating functional for this situation is given by

\be
\label{genfun}
Z=\int \D\bar{\psi}\D\psi \exp{\left\{i\int d^2x \bar{\psi}\left(i\slashed{D}+i\frac{m^2}{2}\frac{\slashed{n}}{n\cdot D}\right)\psi\right\}}.
\ee

\begin{figure}[h!]
\centering
\includegraphics[scale=0.65]{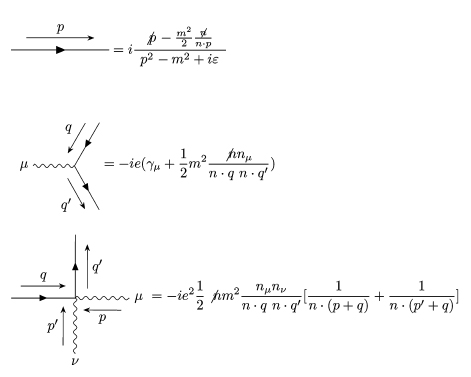}
\caption{Useful Feynman rules in VSR-QED.}
\label{fig:rules}
\end{figure}

From the equation (\ref{genfun}) we get the Feynman rules for VSR-QED. Due to the nonlocal term $(n\cdot D)^{-1}$ there will be an infinite number of vertices in the series. However, if we work in the Light Cone Gauge (LCG) those vertices with two or more photon legs will not contribute. Nevertheless, for the general case, we list in figure \ref{fig:rules} the Feynman rules including the first new vertex.

\begin{figure}[h!]
\centering
\includegraphics[scale=0.6]{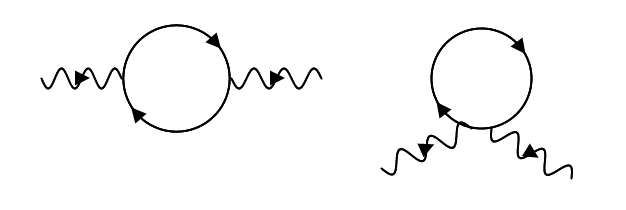}
\caption{Photon vacuum polarization diagrams in VSR.}
\label{fig:photvac}
\end{figure}

To see how it works, we will compute the lowest order vacuum polarization $\Pi^{\mu\nu}$ in two dimensions. We observe a new diagram appears due to the new vertex coming from the non-local operator expansion, which is displayed in the figure \ref{fig:photvac}.\\

As we mentioned before, in the light cone gauge the operator $(n\cdot D)^{-1}$ reduces to $(n\cdot \partial)^{-1}$. Here, we will compute the diagrams without working in a specific gauge to show the result satisfies the Ward identity. In the next section, when we will compute the currents, for the sake of simplicity, we will use the light cone gauge. Therefore, using the Feynman rules in the figure \ref{fig:rules} the vacuum polarization is written as 

\bea
i\Pi_{1 \mu \nu} &=& - e^2 \int \frac{d^2 p}{(2 \pi)^2} \frac{1}{(p^2 - m^2 +
i \varepsilon) ((p - q)^2 - m^2 + i \varepsilon)} tr \left\{ \left(
\gamma_{\mu} + \frac{1}{2} m^2 \frac{\not{n} n_{\mu}}{n \cdot p n \cdot (p -
q)} \right) \left( \slashed{p} - \frac{m^2}{2} \frac{\slashed{n}}{n \cdot p}
\right) \right.\nn\\
& &\left.\times\left( \gamma_{\nu} + \frac{1}{2} m^2 \frac{\slashed{n} n_{\nu}}{n \cdot p
n \cdot (p - q)} \right) \left( \slashed{p} - \slashed{q}  - \frac{m^2}{2}
\frac{\slashed{n}}{n \cdot (p - q)} \right) \right\},\\
i\Pi_{2 \mu \nu} &=& - \frac{1}{2} e^2 m^2 n_{\mu} n_{\nu} \int
\frac{d^2 p}{(2 \pi)^2} \frac{1}{(n \cdot p)^2} \left( \frac{1}{n \cdot (p
+ q)} + \frac{1}{n \cdot (p - q)} \right) \frac{1}{p^2 - m^2 + i
\varepsilon} tr \left\{ \slashed{n} \left( \slashed{p}- \frac{m^2}{2}
\frac{\not{n}}{n \cdot p} \right) \right\}.
\eea

We proceed using dimensional regularization. In order to compute the SIM(2) integrals we use the following decomposition formula
\be
\frac{1}{(n\cdot(p+k_i))(n\cdot(p+k_j))}=\frac{1}{n\cdot(k_i-k_j)} \left(\frac{1}{(n\cdot(p+k_j)}-\frac{1}{n\cdot(p+k_i)} \right),
\ee
Next, we make a change of variables wherever is necessary, and the integrals with $(n\cdot p)^{-1}$ are computed using the Mandelstam-Leibbrandt prescription.  The important formulas were derived in \cite{Alfaro:2016pjw} starting from a symmetry property of $n$ and a new null vector $\bar{n}$ introduced to regulate the integrals. These integrals are listed in the Appendix \ref{app-int}. The introduction of the new null vector $\bar{n}$ breaks the SIM(2) invariance. To recover it we follow the reference \cite{Alfaro:2017umk} trading $\bar{n}$ with a linear combination of $n$ and the external momentum as
\be
\bar{n}_\mu=-\frac{q^2}{2(n\cdot q)^2}n_\mu+\frac{q_\mu}{n\cdot q}.
\ee

After the calculation we get
\be
\label{selfphotongen}
i \Pi_{\mu \nu} = \alpha (q^2) \left[q^2g_{\mu \nu} -q_{\mu} q_{\nu}\right] + \beta
(q^2) \left[- g_{\mu \nu}+\frac{n_{\nu} q_{\mu} + n_{\mu} q_{\nu}}{n \cdot q}
 - q^2\frac{n_{\mu} n_{\nu}}{(n \cdot q)^2}\right],
\ee
where
\bea
\label{alpha0}
\alpha (q^2) &=& - \frac{i e^2}{\pi} \int^1_0 d x \left( \frac{x (1 - x)}{m^2 -
x (1 - x) q^2-i\varepsilon} \right),\\
\label{beta0}
\beta (q^2) &=& \frac{i e^2 m^2}{2 \pi} \int^1_0 d x \frac{x q^2}{(m^2 - x q^2-i\varepsilon)
(m^2 - x (1 - x) q^2-i\varepsilon)}.
\eea

The equation (\ref{selfphotongen}) satisfies the Ward identity $q_\mu\Pi^{\mu\nu}=0$ as it is required by the gauge invariance. From the expressions (\ref{alpha0}) and (\ref{beta0}) we see that $\alpha (0)$ is finite and $\beta (0)=0$. It means in the two dimensional QED-VSR the photon does not get mass, as in the standard QED.\\

When $m^2$ goes to zero in (\ref{alpha0}) and (\ref{beta0}) we recover the standard result. Here, as the VSR result contains the mass term, when we perform the limit $\varepsilon\to0$ we observe in $\alpha(q^2)$ there is a branch cut where $m^2 -x (1 - x) q^2<0$. The product $x(1-x)$ is at most $1/4$. Hence, the branch cut begins at $q^2=4m^2$, which corresponds to the threshold for the creation of an electron-positron pair. So, the theory admits pair production in two dimensions.\\

We will leave the expressions for $\alpha(q^2)$ and $\beta(q^2)$ in this way to move on to the computations of the currents and next, the axial anomaly, where only $\alpha(q^2)$ will be necessary. The integrals are computed in the Appendix \ref{app-int2}.

\section{Vector Current}
Now, we will compute expectation value for the current $j^\mu$. From the generating functional in the equation (\ref{genfun}) we can get the effective action $\Gamma$ using $\Gamma=\log{Z}$. The current is obtained deriving respect to $A_\mu$,
\be
\langle j^\mu\rangle=\frac{\delta \Gamma}{\delta A_\mu}.
\ee

Thus, we have
\be
\langle j^{\mu} \rangle = \frac{1}{Z} \int \mathcal{D} \bar{\psi} \mathcal{D}
\psi \left(  \bar{\psi} \gamma^{\mu} \psi + \frac{1}{2} m^2 \left( \frac{1}{n
\cdot D^{\dag}} \bar{\psi} \right) \slashed{n} n^{\mu} \left( \frac{1}{n \cdot
D} \psi \right) \right) \exp \left[ i \int d^2 x\mathcal{L} \right].
\ee
We will work in the LCG and we get
\be
\langle j^{\mu} \rangle = \frac{1}{Z} \int \mathcal{D} \bar{\psi} \mathcal{D}
\psi \left(  \bar{\psi} \left( \gamma^{\mu} + \frac{m^2}{2} \frac{\slashed{n}
n^{\mu}}{(n \cdot \overset{\leftarrow}{\partial}) (n \cdot \partial)} \right) \psi \right) \exp
\left[ i \int d^2 x\mathcal{L}_0 \right] \exp \left[ - i e \int d^2 x
\bar{\psi} \slashed{A} \psi \right],
\ee
where we recall $\mathcal{L}_0$ is the free fermion lagrangian in the equation (\ref{lagfree}). We proceed perturbatively only to one loop, because following the argument in \cite{Harvey:2005it}, the chiral anomaly is exact at one loop. Although the argument was used in the standard computation, it holds here, since we can interpret the anomaly topologically. Thus, as topological quantities cannot change continuously, perturbative corrections at higher than one loop should not appear. Hence

\bea
\langle j^{\mu} (x) \rangle &=& \lim_{x' \rightarrow x} tr\left[ \left(
\gamma^{\mu} + \frac{1}{2} m^2 \frac{1}{n \cdot \partial_{x'}} \slashed{n}
n^{\mu} \frac{1}{n \cdot \partial_x} \right) S_F (x - x') \right]\nn\\
&-& \lim_{x' \rightarrow x} i e \int d^2 y \quad tr\left[ \left( \gamma^{\mu}
+ \frac{1}{2} m^2 \frac{1}{n \cdot \partial_{x'}} \slashed{n} n^{\mu} \frac{1}{n
\cdot \partial_x} \right) S_F (x - y) \slashed{A} (y) S_F (y - x') \right].\nn\\
\eea

Where, we have replaced the dependence on $x$ in $\bar{\psi}$ with the limit $x'\to x$ to distinguish where the non-local operator $n \cdot \partial$ acts, and we indicate with a subscript in the partial derivatives the variable to be derived. We write the above expression in the Fourier space and after some algebra we get
\be
\langle j^{\mu} (q) \rangle =(- i e) \int \frac{d^2 p}{(2 \pi)^2} tr \left[ \left( \gamma^{\mu} + \frac{1}{2} m^2 \frac{\slashed{n} n^{\mu}}{(n \cdot (p -
q)) (n \cdot p)} \right) \frac{i}{\slashed{p} - \frac{m^2}{2}
\frac{\slashed{n}}{n \cdot p}} \left( \gamma^{\nu} \right)\frac{i}{\slashed{p} - \slashed{q} - \frac{m^2}{2} \frac{\slashed{n}}{n \cdot (p - q)}}
\right] A_{\nu} (q).
\ee

This object is the same vacuum polarization after using the condition $n\cdot A=0$, except an $i/e$ factor. Thus, the expectation value for the current is given by

\be
\label{currentexpval}
\langle j^{\mu} (q) \rangle = \frac{i}{e}\alpha (q^2) \left[q^2 A^{\mu} -q^{\mu} q\cdot A\right] +  \frac{i}{e}\beta(q^2) \left[- A^{\mu}+\frac{n^{\mu} q\cdot A}{n \cdot q}\right].
\ee

We notice the expectation value of the current is conserved, $q_\mu\langle j^{\mu} (q) \rangle=0$.

\section{Axial Anomaly}
We will compute the expectation value for $j^{\mu5}$. We use the equation (\ref{jrelation2}), which relates $j^\mu$ with $j^{\mu5}$, and we use the equation (\ref{currentexpval}) to get

\be
\label{jmu5pre}
j^{\mu5}=-\frac{i}{e}\alpha (q^2) \epsilon^{\mu\nu}\left[q^2 A_{\nu} -q_{\nu} q\cdot A\right] -  \frac{i}{e}\beta(q^2) \epsilon^{\mu\nu}\left[- A_{\nu}+\frac{n_{\nu} q\cdot A}{n \cdot q}\right].
\ee

We contract (\ref{jmu5pre}) with $q_\mu$ and we write it in terms of $F_{\mu\nu}$. Therefore,

\be
q_{\mu} \langle j^{\mu 5} \rangle = - \frac{i}{e} \left[ (\alpha (q^2)q^2 -
\beta (q^2)) \frac{1}{2} \varepsilon^{\mu \nu} F_{\mu \nu} (q) - \beta (q^2)
\frac{n^{\alpha} q^{\beta} F_{\alpha \beta}}{(n \cdot q)^2} \epsilon^{\mu
\nu} q_{\mu} n_{\nu} \right].
\ee

Seemingly, a new anomaly term appears. Nevertheless, since we are working in two dimensions is easy to see that $n^{\alpha} q^{\beta} F_{\alpha \beta}=\frac{1}{2} \varepsilon^{\mu \nu} n_{\mu} q_{\nu} \varepsilon^{\alpha \beta}F_{\alpha \beta}$. Thus,

\be
q_{\mu} \langle j^{\mu 5} \rangle = - \frac{i}{e} \left[ (\alpha (q^2)q^2 -
\beta (q^2)) \frac{1}{2} \varepsilon^{\mu \nu} F_{\mu \nu} (q) + \beta (q^2)
\frac{(\varepsilon^{\alpha \beta} n_{\alpha} q_{\beta})^2}{2 (n \cdot q)^2}
\varepsilon^{\mu \nu} F_{\mu \nu} \right].
\ee
In addition, using $n_0=-n_1=1$ we get as result
\be
\frac{\varepsilon^{\alpha \beta} n_{\alpha} q_{\beta}}{n \cdot q}=1,
\ee
Therefore, the terms with $\beta(q^2)$ cancel out and we get
\be
q_{\mu} \langle j^{\mu 5} \rangle =  - \frac{i}{2e}\alpha(q^2)q^2 \varepsilon^{\mu \nu} F_{\mu \nu} (q),
\ee
We use the result for $\alpha(q^2)$ in the appendix \ref{app-int2}  and finally
\be
\label{anom}
q_{\mu} \langle j^{\mu 5} \rangle =\left[\frac{e}{2\pi} + \frac{e m^2}{\pi q^2 \sqrt{1
- \frac{4 m^2-i\varepsilon}{q^2}}} \log \left( \frac{1 + \sqrt{\frac{q^2 - 4 m^2 + i
\varepsilon}{q^2}}}{- 1 + \sqrt{\frac{q^2 - 4 m^2 + i \varepsilon}{q^2}}}
\right)\right]\varepsilon^{\mu \nu} F_{\mu \nu} (q).
\ee

In the limit $m\to0$, we recover the standard result. In that case, the equation (\ref{anom}) reduces to the result displayed in the equation (\ref{stanomaly}). In the limit $q^2\to0$ and $\varepsilon\to0$ the result is zero. So, the anomaly vanishes. The standard result reached in the limit $m\to0$ is equivalent to take $q\to\infty$. Thus, our result interpolates between large momentum (short distances) where an anomaly is appreciated and the low momentum (large distances) where there is not anomaly. Furthermore, we see in the equation (\ref{anom}) that the same topological invariant appears as in the standard computation. However, there is a modification in the coefficient next to the anomaly term due to the VSR term.\\

\section{Conclusions}

The existence of the null vector $n=(1,1)$ that transforms by a phase in the two dimensional Lorentz group allowed us to study the Schwinger model under the VSR framework. Terms which contains $n$ in fractions could be incorporated elsewhere.\\

We have found in section V that the photon does not receive a mass since $\alpha (0)$ is finite and $\beta (0)=0$. This result is completely different from the Schwinger work in the standard $QED_2$, where the photon acquires a mass $e^2/\pi$. In addition, from the vacuum polarization computation, we observe pair production in this model.\\



Both the free vector current and the free axial current change when we couple the fermion with only a VSR mass to an external electromagnetic field due to the non-local operator $(n\cdot D)^{-1}$. However, it reduces to the free case in the light cone gauge $n\cdot A=0$. Since the VSR current is different from the standard case, our calculation of the axial anomaly presents a difference in the coefficient which accompanies the anomaly term respect to the standard result.\\

\begin{acknowledgements}

The work of A. Soto is supported by the CONICYT-PFCHA/Doctorado Nacional/2017-21171194 and
Fondecyt 1150390. The work of J. Alfaro is partially supported by Fondecyt 1150390 and CONICYT-PIA-ACT1417.\\

\end{acknowledgements}
\vspace{-20pt}

\begin{appendix}

\section{Integration with $(n\cdot p)^{-1}$} \label{app-int}
In this appendix we list the main integrals needed to compute our expressions with $(n\cdot p)^{-1}$ in the paper, which are obtained from ref.\cite{Alfaro:2016pjw}. They are
\be
\label{ap1}
\int dp\frac{1}{(p^2+2p\cdot q-m^2)^a}\frac{1}{(n\cdot p)^b}=(-1)^{a+b}i\pi^\omega(-2)^b\frac{\Gamma(a+b)}{\Gamma(a)\Gamma(b)}(\bar{n}\cdot q)^b\int_0^1 dt t^{b-1}\frac{1}{(m^2+q^2-2(n\cdot q)(\bar{n}\cdot q)t)^{a+b-\omega}},
\ee
with $\omega=d/2$.\\

Taking a derivative with respect to $q$ in (\ref{ap1}) we get
\bea
\int d^d p \frac{p_{\mu}}{(p^2 + 2 p \cdot q - m^2)^{a + 1}} \frac{1}{(n
 \cdot p)^b} &=&
(- 1)^{a + b} i \pi^{\omega} (- 2)^{b - 1} \frac{\Gamma (a + b -
\omega)}{\Gamma (a + 1) \Gamma (b)} (\bar{n} \cdot q)^{b - 1} b
\bar{n}_{\mu} \int^1_0 d t t^{b - 1} \frac{1}{[m^2 + q^2 - 2 (n  \cdot q)
(\bar{n} \cdot q) t]^{a + b - \omega}}\nn\\
&+&(- 1)^{a + b} i \pi^{\omega} (- 2)^b \frac{\Gamma (a + b + 1 -
\omega)}{\Gamma (a + 1) \Gamma (b)} (\bar{n}  \cdot q)^b \int^1_0 d t t^{b -
1} \frac{q_{\mu} - t (n  \cdot q \bar{n}_{\mu} + \bar{n}  \cdot q
n_{\mu})}{[m^2 + q^2 - 2 (n  \cdot q) (\bar{n}  \cdot q) t]^{a + b + 1 -
\omega}}.\nn\\
\eea

Taking a second derivative the integral is 

\bea
\int d^d p \frac{p_{\mu} p_{\nu}}{(p^2 + 2 p \cdot q - m^2)^{a + 2}}
\frac{1}{(n \cdot p)^b} &=&(- 1)^{a + b} i \pi^{\omega} (- 2)^{b - 2}\times\nn\\
&&\times\left\{ \frac{\Gamma (a + b -
\omega)}{\Gamma (a + 2) \Gamma (b - 1)} (\bar{n} \cdot q)^{b - 2} b
\bar{n}_{\mu} \bar{n}_{\nu} \int^1_0 d t t^{b - 1} \frac{1}{(m^2 + q^2 - 2 (n
\cdot q) (\bar{n} \cdot q) t)^{a + b - \omega}}\right.\nn\\
&&\left.- 2 \frac{\Gamma (a + b +1 - \omega)}{\Gamma (a + 2) \Gamma (b)} (\bar{n} \cdot q)^{b - 1} b
\bar{n}_{\mu} \int^1_0 d t t^{b - 1} \frac{q_{\nu} - t (n \cdot q
\bar{n}_{\nu} + \bar{n} \cdot q n_{\nu})}{(m^2 + q^2 - 2 (n \cdot q)
(\bar{n} \cdot q) t)^{a + b + 1 - \omega}}\right.\nn\\
&&\left. - 2 \frac{\Gamma (a + b + 1 -
\omega)}{\Gamma (a + 2) \Gamma (b)} \left( \bar{n} \cdot q \right)^{b - 1} b
\bar{n}_{\nu} \int^1_0 d t t^{b - 1} \frac{q_{\mu} - t (n \cdot q
\bar{n}_{\mu} + \bar{n} \cdot q n_{\mu})}{(m^2 + q^2 - 2 (n \cdot q)
(\bar{n} \cdot q) t)^{a + b + 1 - \omega}} \right.\nn\\
&&\left.+ 4 \frac{\Gamma (a + b + 2 -
\omega)}{\Gamma (a + 2) \Gamma (b)} (\bar{n} \cdot q)^b \int^1_0 d t t^{b -
1} \frac{[q_{\nu} - t (n \cdot q \bar{n}_{\nu} + \bar{n} \cdot q n_{\nu})]
[q_{\mu} - t (n \cdot q \bar{n}_{\mu} + \bar{n} \cdot q n_{\mu})]}{(m^2 +
q^2 - 2 (n \cdot q) (\bar{n} \cdot q) t)^{a + b + 2 - \omega}} \right.\nn\\
&&\left.- 2
\frac{\Gamma (a + b + 1 - \omega)}{\Gamma (a + 2) \Gamma (b)} (\bar{n} \cdot
q)^b \int^1_0 d t t^{b - 1} \frac{g_{\mu \nu} - t (n_{\nu} \bar{n}_{\mu} +
\bar{n}_{\nu} n_{\mu})}{(m^2 + q^2 - 2 (n \cdot q) (\bar{n} \cdot q) t)^{a
+ b + 1 - \omega}} \right\}
\eea

\section{Solution of the integrals in the vacuum polarization} \label{app-int2}
Here we will compute $\alpha(q^2)$ and $\beta(q^2)$. We recall the expressions (\ref{alpha0}) and (\ref{beta0})
\bea
\alpha (q^2) &=& - \frac{i e^2}{\pi} \int^1_0 d x \left( \frac{x (1 - x)}{m^2 -
x (1 - x) q^2-i\varepsilon} \right),\\
\beta (q^2) &=& \frac{i e^2 m^2}{2 \pi} \int^1_0 d x \frac{x q^2}{(m^2 - x q^2-i\varepsilon)
(m^2 - x (1 - x) q^2-i\varepsilon)}.
\eea


To simplify the computation we split in partial fractions as $\alpha(q^2)$ as $\beta(q^2)$. To do this in $\beta(q^2)$ is essential to keep $m^2\neq 0$. After this, we have
\bea
\alpha (q^2) &=& \frac{i e^2}{\pi q^2} + \frac{i e^2 m^2}{\pi (q^2)^2 \sqrt{1 - 
\frac{4m^2-i\varepsilon}{q^2}}} (I_1-I_2),\\
\beta (q^2) &=& \frac{i e^2}{4 \pi} \left( \frac{1}{\sqrt{1 - 
\frac{4m^2-i\varepsilon}{q^2}}} + 1 \right) I_1 - \frac{i e^2}{4 \pi} \left(
\frac{1}{\sqrt{1 - \frac{4m^2-i\varepsilon}{q^2}}} - 1 \right) I_2 - \frac{i e^2}{2 \pi}I_3,
\eea

where the terms $I_1$, $I_2$ and $I_3$ are defined as
\bea
I_1 &=& \int^1_0 d x \frac{1}{x - \frac{1}{2} \left( 1 - \sqrt{1 -
\frac{4m^2-i\varepsilon}{q^2}} \right)},\\
I_2 &=& \int^1_0 d x \frac{1}{x - \frac{1}{2} \left( 1 + \sqrt{1 - 
\frac{4m^2-i\varepsilon}{q^2}} \right)},\\
I_3 &=& \int^1_0 d x \frac{1}{x - \frac{m^2-i\varepsilon}{q^2}}.
\eea


We solve the integrals and we get
\bea
I_1 &=& \log\left(\frac{1+\sqrt{\frac{q^2-4m^2+i\varepsilon}{q^2}}}{-1+\sqrt{\frac{q^2-4m^2+i\varepsilon}{q^2}}}\right),\\
I_2 &=& \log\left(\frac{-1+\sqrt{\frac{q^2-4m^2+i\varepsilon}{q^2}}}{1+\sqrt{\frac{q^2-4m^2+i\varepsilon}{q^2}}}\right),\\
I_3 &=& \log(q^2-m^2+i\varepsilon)-\log(-m^2+i\varepsilon).
\eea


and the values of $\alpha(q^2)$ and $\beta(q^2)$ are
\bea
\alpha (q^2) &=& \frac{i e^2}{\pi q^2} + \frac{2 i e^2 m^2}{\pi (q^2)^2 \sqrt{1
- \frac{4 m^2-i\varepsilon}{q^2}}} \log \left( \frac{1 + \sqrt{\frac{q^2 - 4 m^2 + i
\varepsilon}{q^2}}}{- 1 + \sqrt{\frac{q^2 - 4 m^2 + i \varepsilon}{q^2}}}
\right),\\
\beta (q^2) &=& - \frac{i e^2}{2 \pi} [\log (q^2 - m^2 + i \varepsilon) - \log
(- m^2 + i \varepsilon)] + \frac{i e^2}{2 \pi \sqrt{1 - \frac{4 m^2}{q^2}}}
\log \left( \frac{1 + \sqrt{\frac{q^2 - 4 m^2 + i \varepsilon}{q^2}}}{- 1 +
\sqrt{\frac{q^2 - 4 m^2 + i \varepsilon}{q^2}}} \right).
\eea




\end{appendix}

\end{document}